\documentclass[atoms,article,accept,pdftex,moreauthors]{Definitions/mdpi} 

\firstpage{1} 
\pubvolume{11}
\issuenum{2}
\articlenumber{30}
\pubyear{2023}
\copyrightyear{2023}
 \externaleditor{Academic Editors: { Alexander Kramida and Ulrich D. Jentschura}}
 
\datereceived{31 December 2022} 
\daterevised{1 February 2023} 
\dateaccepted{2 February 2023} 
\datepublished{5 February 2023} 
\hreflink{https://\linebreak doi.org/10.3390/atoms11020030} 

\Title{Deep Laser Cooling of Thulium Atoms to Sub-\textmu K Temperatures in Magneto-Optical Trap}

\TitleCitation{Deep Laser Cooling of Thulium Atoms to Sub-\textmu K Temperatures in Magneto-Optical Trap}

\usepackage{braket}
 \usepackage{trackchanges}

\newcommand{\unit}[1]{\,\textrm{#1}}
\newcommand{\red}[1]{\textcolor{red}{#1}}

\Author{Daniil 
 Provorchenko $^{1,}$*\orcidA{}, Dmitry Tregubov $^{1,}$*\orcidB{}, Denis Mishin $^{1}$, Mikhail Yaushev $^{1}$, Denis Kryuchkov $^{1}$, Vadim~Sorokin $^{1}$, Ksenia Khabarova $^{1,2}$, Artem Golovizin $^{1}$\orcidC{} and Nikolay Kolachevsky  $^{1,2}$}

\AuthorNames{Daniil Provorchenko, Dmitry Tregubov, Denis Mishin, Artem Golovizin, Mikhail Yaushev, Denis Kryuchkov, Vadim Sorokin, Ksenia Khabarova and Nikolay Kolachevsky}

\AuthorCitation{Provorchenko, D.; Tregubov, D.; Mishin, D.;  Yaushev, M.; Kryuchkov, D.; Sorokin, V.; Khabarova, K.; Golovizin, A.;  Kolachevsky, N.}

\address{%
$^{1}$ \quad The Lebedev Physical Institute of the Russian Academy of Sciences, Moscow 119991, Russia 
$^{2}$ \quad Russian Quantum Center, Moscow 121205, Russia
}

\corres{Correspondence: provorchenko.di@phystech.edu (D.P.); treg.dim@gmail.com (D.T.)}

\abstract{
Deep laser cooling of atoms, ions, and molecules facilitates the study of fundamental physics as well as applied research.
In this work, we report on the narrow-line laser cooling of thulium atoms at the wavelength of $506.2 \unit{nm}$ with the natural linewidth of $7.8\unit{kHz}$, which widens the limits of atomic cloud parameters control. 
Temperatures of about $400 \unit{nK}$, phase-space density of up to $3.5\times10^{-4}$ and $2\times10^6$ number of trapped atoms were achieved.
We have also demonstrated formation of double cloud structure in an optical lattice by adjusting parameters of the  $506.2 \unit{nm}$ magneto-optical trap.
These results can be used to improve experiments with BEC, atomic interferometers, and optical~clocks.
}

\begin{document}
\section{Introduction}
\label{sec:intro}

In modern atomic physics, laser cooling of atoms, ions, and molecules is an essential technique for producing cold and dense clouds of particles. Such ensembles are the subject of study in many fields, including few- and many-body physics \cite{blume2012few, bloch2012quantum}, {absolute and differential quantum gravimetry} \cite{janvier2022compact, Menoret2018, DelAguila2018Bragg88Sr}, quantum simulations \cite{bloch2012quantum}, Bose--Einstein condensate (BEC) \cite{ketterle1996bose, warner2021overlapping, miyazawa2022bose, schreck2021laser, davletov2020machine}, and optical lattice clocks \cite{ludlow2015optical, boulder2021frequency}.  

Bose--Einstein condensation requires the atomic de Brogile wavelength to be compared with atom--atom distance.
The main aim of laser cooling techniques in such experiments is to achieve a better starting point for evaporative cooling. 
Laser cooling techniques typically allows to reach the temperatures of about ($\sim$$1\unit{\textmu K}$) with the typical phase-space density (PSD) of $10^{-4}$ \cite{griesmaier2005bose, lu2011strongly, kraft2009bose, aikawa2012bose, stellmer2009bose}, {while the PSD of 2.612 is} required to achieve BEC \cite{ketterle1996bose}.
A further increase in PSD is achieved with evaporative cooling which reduces the number of particles.

Another application of laser-cooled ensembles is the optical {lattice} clocks. 
The state-of-the-art level of relative frequency uncertainty of lattice optical clocks is on a scale of $10^{-18}$ \cite{ludlow2015optical}. 
In such experiments, cold atoms are trapped in optical lattices at the magic wavelengths to reduce the lattice-induced frequency shifts {of the clock transition}, but high-order polarizabilities lead these frequency shifts to be dependent on vibrational states of atoms in the potential 
\cite{Ushijima2018operational}.
High temperature means that more vibrational states are occupied, leading to undesirable frequency shifts and to decoherence of the clock transition excitation \cite{Blatt2009, Fedorova2020RabiLattice}. 
It is desirable that a significant number of atoms should be in the lowest vibrational state, which usually corresponds to temperatures of the order of ($\sim$$1\unit{\textmu K}$). 

Thulium atoms are proven to be a convenient platform for optical clocks \cite{Golovizin2019}, quantum simulators, and experiments with BEC \cite{KhlebnikovRandomAtoms,davletov2020machine}.
The clock transition in thulium has low sensitivity to most environmental frequency shifts while others can be eliminated by the interrogation protocol \cite{Golovizin2021}.
In the case of quantum simulators and BEC experiments, thulium atoms are promising due to large number of Fano--Feshbach resonances and relatively high magnetic moment of the ground state ($\mu  = 4 \mu _B$) \cite{khlebnikov2021characterizing}. 

All current experiments with cold atomic thulium {are performed with the single stable bosonic isotope thulium-169. 
The experimental procedure} starts with two-stage laser cooling which typically produces several millions of atoms at a temperature of $\sim$$20\unit{\textmu K}$ . 
In experiments with BEC, this results in PSD of $\sim$$2 \times 10^{-5}$ after the second stage cooling \cite{davletov2020machine}.
In optical clock experiments, such temperature leads to mean vibrational number $n_\textrm{vib} \approx 8$, which corresponds to the relative frequency shift at the $10^{-16}$ level and hard-to-control uncertainty \cite{mishin2022effect}.

\begin{figure}[H]
\begin{adjustwidth}{-\extralength}{0cm}
\centering
\includegraphics[width=\textwidth]{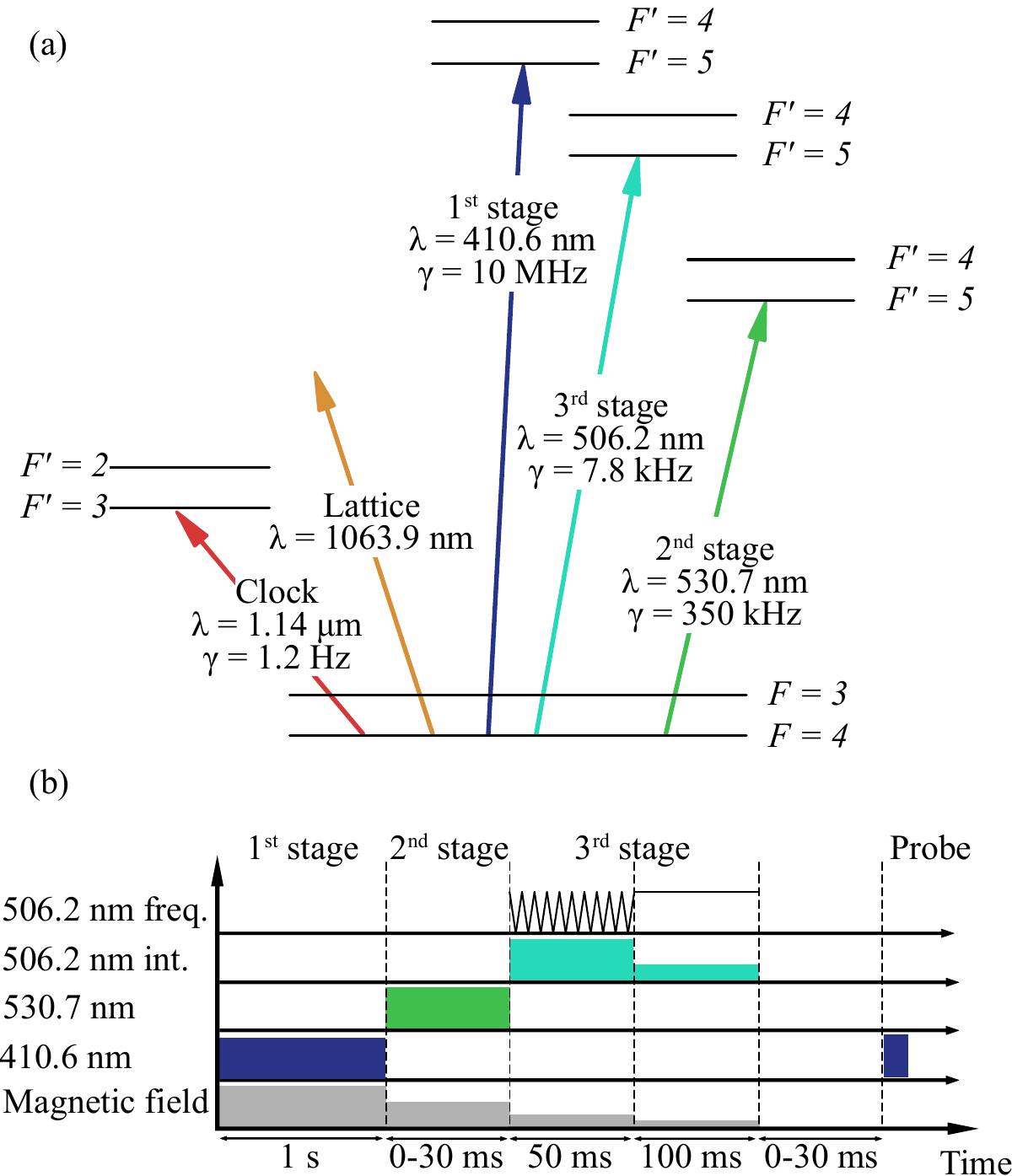}
\end{adjustwidth}
\caption{
(\textbf{a}) The 
 diagram of levels and transitions in Tm that we used in this work. 
For the 1{st}-stage cooling and the probe beam we use the radiation at the wavelength of $410.6\unit{nm}$. 
For the 2{nd}-stage cooling---$530.7\unit{nm}$, and the 3{rd}-stage cooling---$506.2\unit{nm}$, the transition that we study in this work.
The radiation at $1063.9\unit{nm}$ used to form the optical lattice and the clock transition at $1.14\unit{\textmu m}$ used to confirm the low temperatures, achieved in experiment.
(\textbf{b}) The typical pulse sequence and its timings used in our experiments. 
The diagram shows the order of laser cooling pulses and the magnetic field changes with time.
For 3rd stage cooling pulse the frequency and the intensity of cooling radiation showed independently.
\label{fig:scheme}}
\end{figure}

Thus, a deeper cooling of thulium atoms is preferable to solve both of these problems.
The minimal temperature $T_\textrm{D}$ attainable with Doppler cooling is related to the linewidth of cooling transition $\gamma$ according to the relationship $T_\textrm{D} = h \gamma /(2k_\textrm{B})$, where $h$---the Planck constant, and $k_\textrm{B}$---the Boltzmann constant.
The first transition used for laser cooling of Tm atoms is {the transition at the wavelength of} $410.6\unit{nm}$ (``blue'', 1{st} stage magneto-optical trap, MOT, {Figure}~\ref{fig:scheme}a) with the natural linewidth of $10\unit{MHz}$. 
While Doppler limit for this transition equals $240\unit{\textmu K}$, it is possible to work in a sub-Doppler regime with temperatures as low as $25\unit{\textmu K}$. 
However, the atomic cloud in this case becomes larger with lower number of atoms.
The second laser cooling transition has the wavelength of $530.7\unit{nm}$ (``green'', 2{nd} stage MOT, {Figure} \ref{fig:scheme}a) and the linewidth of $350\unit{kHz}$. 
The recapture efficiency from the first stage MOT is close to 100\%, and the Doppler limit equals $10\unit{\textmu K}$.

In this work, we present a detailed experimental analysis of not yet studied transition $\ket{4f^{13}(^2F^o)6s^2,\, F = 4} \xrightarrow{} \ket{4f^{13}(^2F_{7/2}^o)6s6p(^3P_2^o),\, F = 5}$ for deep cooling of thulium atoms, at the wavelength of $\lambda = 506.2\unit{nm}$ with the linewidth of $\gamma = 7.8\unit{kHz}$. 
The Doppler limit for this transition is $T_\textrm{D} = 200\unit{nK}$ which is lower than the recoil limit for this transition $T_\textrm{rec} = h^2/(2m\lambda^2k_\textrm{B}) = 220\unit{nK}$, where $m$---atomic mass of thulium.

{In Section} \ref{sec:results} {we describe our experimental results:
the efficient recapture of atoms from the} 1{st} and the 2{nd} into the 3{rd}-stage MOT (Section \ref{sec:broad}),
{the optimization of the narrow line cooling process} (Section~\ref{sec:search}),
{the confirmation of the low temperature of MOT with the Doppler broadening of the clock transition spectroscopy} (Section~\ref{sec:clock}), {the effect of the intermediate} 2{nd}-stage MOT (Section~\ref{sec:2nd}), {and the possibility to form double-structured clouds in an optical lattice} (Section~\ref{sec:double}).
{Finally, we draw the conclusions in Section} \ref{sec:discuss}.

\section{Results}
\label{sec:results}

\subsection{Loading of Broadband MOT}
\label{sec:broad}
As a starting point, we cool and trap thulium atoms using two-stage MOT (see Figure~\ref{fig:scheme}b).
The Zeeman slower and the first-stage MOT work on the strong transition at a wavelength of $410.6\unit{nm}$ with the natural linewidth of $10\unit{MHz}$.
{This transition is also used for the probe radiation.}
The second-stage MOT works on $530.7\unit{nm}$ transition with natural linewidth of $350\unit{kHz}$.
The transfer efficiency of atoms from the 1{st} to 2{nd} stage MOT is near $100 \%$, and the final temperature of atoms is about $20\unit{\textmu K}$.
More details can be found in~\cite{Sukachev2014SecondaryTraps,Kalganova2017Two-temperatureTrap}.

{In this work}, we follow a common approach of working with narrow-line magneto-optical trap \cite{katori1999magneto, kraft2009bose, yamaguchi2019narrow}, when in the beginning {of narrow-line cooling} the cooling radiation frequency is modulated {to broaden its spectrum and increase the velocity capturing range}.
Note that the laser itself is stabilized to an ultrastable cavity (see Appendix \ref{sec:append}).
We use triangle-shaped frequency modulation for convenience, as it was readily available for our direct digital synthesizers (DDS), based on AD9910 chip. 
The spectrum of triangle-shaped modulated signal differs only slightly from the spectrum of a more commonly used harmonic frequency modulation.

We determined modulation parameters that work well for our setup: amplitude of $100\unit{kHz} \approx 13\gamma$ and frequency of about $10\unit{kHz}$.
{The center frequency of the modulated radiation is typically red-detuned further from the line center than the frequency of the narrow-line cooling stage radiation} (see Figure \ref{fig:scheme}b). 
{Note that there is a wide range of acceptable modulation parameters which may depend on magnetic fields and the atomic cloud geometry and temperature.}

In our case, since we start with much lower temperature of atoms ($\sim$$20\unit{\textmu K}$ vs. \mbox{0.5--3~{mK}}  in \cite{katori1999magneto,lu2011strongly, maier2014narrow}), the $50\unit{ms}$-long broadband cooling results in stable trapping of $\approx$$100\%$ atoms from the 2{nd} stage MOT and $1.5\unit{\textmu K}$ temperature of the atoms.
We use it as a starting point for investigation of single-frequency MOT performance.

\subsection{Optimal MOT Parameters}
\label{sec:search}

The key parameters of atomic cloud after laser cooling are its temperature, size, number of atoms, and density (or phase-space density). 
We measured all of them for different gradients of anti-Helmholtz magnetic field, cooling radiation intensity (in units of saturation parameter $s = I/I_\text{sat}$, where $I_\text{sat} \approx 8 \unit{\textmu W/cm}^2$), and frequency detuning from resonance $\delta\nu$ in units of natural linewidth $\gamma$. 
We varied optical power from $s = 14$ ($P_\textrm{min} = 100 \unit{\textmu W}$, the minimum optical power we could see MOT with) to $s = 89$ ($P_\textrm{max} \approx 635 \unit{\textmu W}$, increasing the optical power further would not lead to increasing the number of atoms or decreasing the temperature).
The detuning was varied from $-8\gamma$ ($-60 \unit{kHz}$, closer to the resonance we lost a significant amount of atoms) to $-14\gamma$ ($-110\unit{kHz}$, {further from the resonance the horizontal cloud size becomes too big and complicates the temperature measurements}).
For the discussion on the line center position, see Appendix \ref{sec:append}.
The magnetic field gradient was varied from $0.2 \unit{G/cm}$ to $0.6\unit{G/cm}$ ($20\unit{mA}$ to $60 \unit{mA}$ for the current in coils). 

{
The pulses timing sequence we used for this experiment is presented in Figure} \ref{fig:scheme}b. 
The duration of 2{nd}-stage cooling was fixed to $30\unit{ms}$. 
The delay time between narrow-line cooling and the probe pulse was varied between 0--30 {ms} for the time-of-flight measurement. 
{We calibrated our CMOS-camera in advance to get the number of atoms as well as the horizontal and vertical cloud dimensions from the images.
The cloud size dependance on the time-of-flight was used to deduce the temperature, and the cloud size before the release of atoms with their number and temperature was used for the PSD calculation.}

\begin{figure}[H]
 
\includegraphics[width=0.98\textwidth]{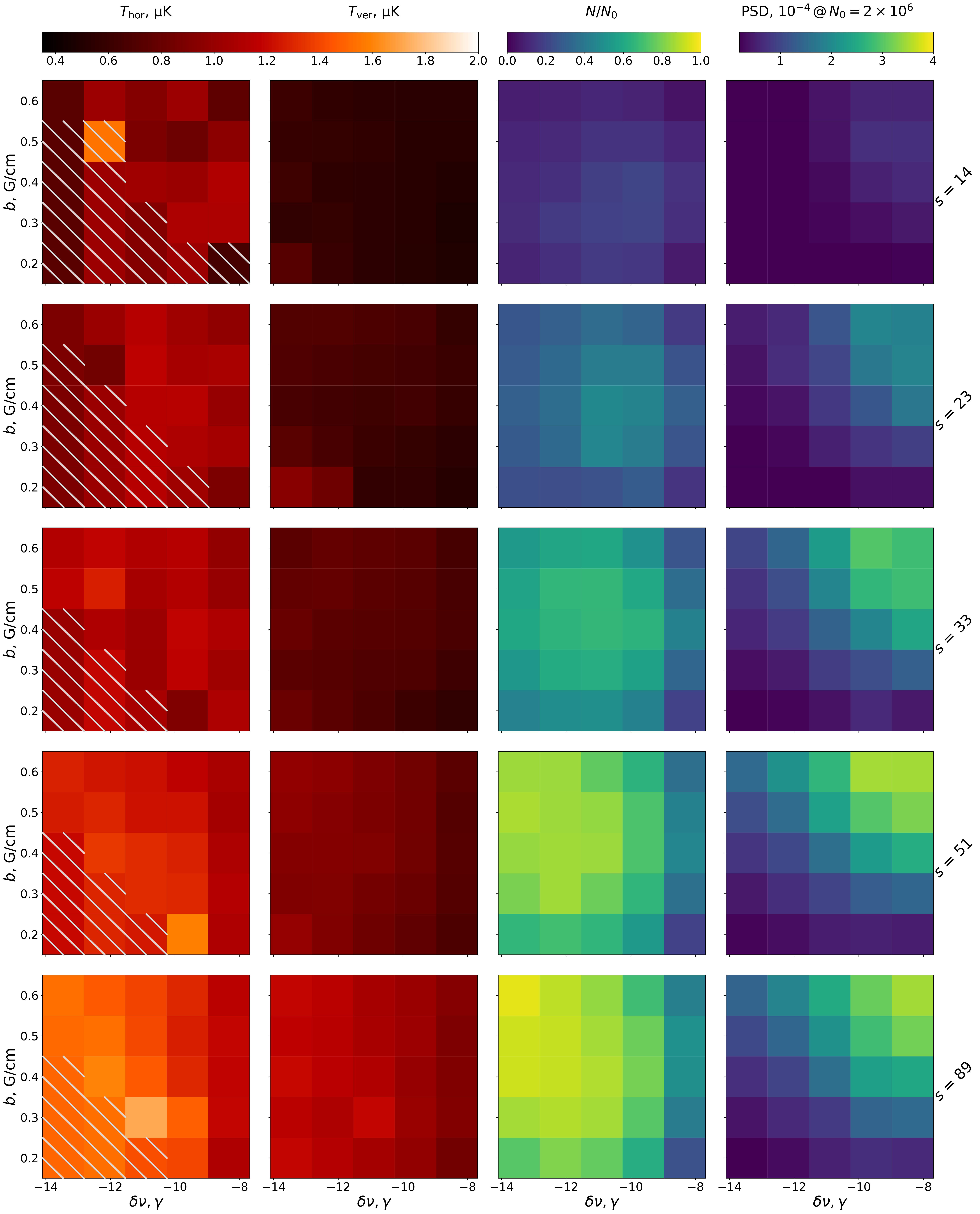}
 
\caption{
 Results of MOT optimization process for several input parameters: magnetic field gradient ($b$, $y$-axis), frequency detuning ($\delta \nu$, $x$-axis), and total intensity of $506.2\unit{nm}$ MOT beams in the units of saturation parameter ($s$, rows).
 Four columns represent four key parameters: two temperatures of atomic cloud ($T_\textrm{hor}$ along the horizontal plane and $T_\textrm{ver}$ along the vertical direction), phase-space density ($PSD$), and relative number of atoms ($N / N_0$, where $N_0 =  2 \times 10^6$---the number of atoms in MOT after first stage cooling). Bowl-shaped cloud in the region of high detunings and low magnetic field gradient complicates temperature measurement in the horizontal plane. 
 This area gives unreliable results for the horizontal temperature and is marked by the white lines crossing this~area.
 \label{fig:optimization}}
\end{figure}

The results are shown in Figure\,\ref{fig:optimization}.
We show atomic temperatures in the horizontal and vertical directions (measured by time-of-flight {technique}), fraction of {recaputered} atoms $N/N_0$, and phase-space density after $100\unit{ms}$ of single-frequency MOT.

Firstly, one can see that minimum temperature is achieved at low intensities and small detunings. 
However, in this parameter space the number of trapped atoms is less than $20 \%$, and the PSD is below $10^{-4}$.
The fraction of trapped atoms can be significantly increased almost up to $100 \%$ by increasing intensity, detuning, and magnetic field gradient. 
The explanation is that the velocity capturing range is increasing.
Maximum PSD of $3.5\times10^{-4}$ is achieved at $s=51$,  $b=0.6\unit{G/cm}$, and $\delta\nu=-10\gamma$ and corresponds to a compromise between the cloud temperature ($T_\text{ver}=0.8\unit{\textmu K}$, $T_\text{hor}=1.2\unit{\textmu K}$) and the number of atoms ($N=1.3\times10^6$ or $N/N_0=64\%$), and is shifted to low detunings which makes the atomic cloud more compact. 
This is a significant order of magnitude increase in PSD compared to the previously reported values for $530.7\unit{nm}$ MOT \cite{davletov2020machine}.




\subsection{Temperature Measurement with Clock Transition Spectroscopy}
\label{sec:clock}
The standard method for temperature measurements in this kind of experiment is time-of-flight method \cite{lett1988observation}.
In order to measure low temperatures, it is preferable to allow for a long time-of-flight of about  20--30~{ms}  which corresponds to a free-fall distance of  2--4~{mm}. 
The field of view of our camera was about $6\unit{mm}$. 
To keep the intensity of the probe beam constant along this distance, we used vertical probe beam. 

Note that for narrow-line Doppler cooling, we often get a typical bowl-shape of the cloud, with horizontal dimensions being significantly larger than vertical. 
In this case we see that laser cooling works more efficiently in the vertical direction than in horizontal plane, and the resulting temperature is inhomogeneous which can be seen in the first two columns in Figure\,\ref{fig:optimization}.

{The cloud temperature in the vertical direction is of particular interest to us because of our experiments with optical lattice clock. 
In such experiments, the clock transition spectroscopy is often performed with an optical lattice aligned vertically.
As mentioned previously in Section} 
\ref{sec:intro}, 
{lowering the vertical temperature of the cloud would reduce the mean longitudinal vibrational sublevel population:}


\begin{equation}
\label{n_z}
    \overline{n_{z}} = \frac{\sum_{n_z=0}^{n_{z}^{\text{max}}} n_z \exp{(-h \nu_z n_z/k_\text{B} T_\text{ver})}}{\sum_{n_z=0}^{n_{z}^{\text{max}}}\exp{(-h \nu_z n_z/k_\text{B} T_\text{ver})}}
\end{equation}

{In this equation,} $n_{z}^{\text{max}}$
is the maximum vibrational number in the potential which is defined by the potential depth, $\nu_z$ is the longitudinal frequency of the optical lattice.

 {In our experiments, we operate the optical lattice at a wavelength of 1063.9 nm which is close to the magic wavelength which we will use in clock spectroscopy experiments}~\mbox{\cite{Golovizin2019, mishin2022effect}}.
The typical vibrational frequency $\nu_z = 40 \unit{kHz}$ and maximum vibrational number \mbox{$n_z^\text{max} = 15$}.
With lowering the temperature of atoms from $20\unit{\textmu K}$ which we achieved in our previous works, to about $1\unit{\textmu K}$ achieved in this work, the mean longitudinal vibrational sublevel population changes from $\overline{n}_z \approx 4.4$ to $\overline{n}_z \approx 0.2$, which is a great starting point for further experiments. 
Note that further decreasing of mean vibrational number could be achieved via optical lattice sifting  (for more information about this method and typical optical lattice parameters in our experiments, see \cite{mishin2022effect}).

\begin{figure}[H]
 
\includegraphics[width=13.5cm]{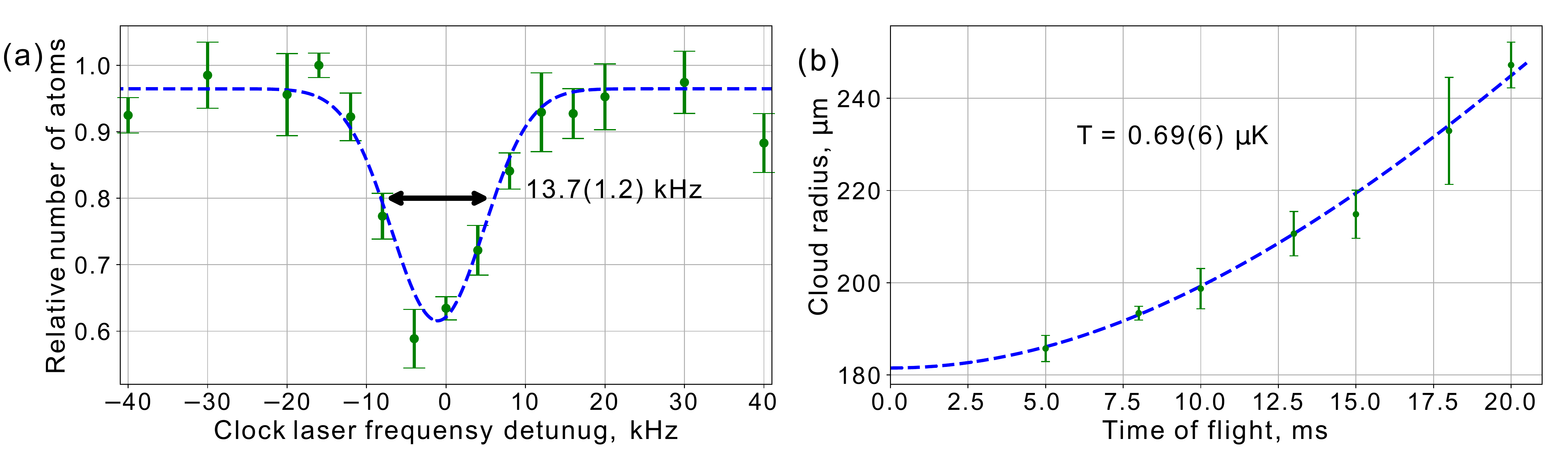}
 
\caption{
(\textbf{a}) Clock 
 line spectroscopy in free-space using vertical clock laser beam: relative number of atoms remained after clock transition excitation depending on clock frequency detuning from resonance.
(\textbf{b}) Atomic cloud vertical size (radius at $1/\textrm{e}$ level) during the time-of-flight experiment. The temperature $T_\textrm{ver} = 0.69(6)\unit{\textmu K}$ explains $12\unit{kHz}$ of the linewidth which is consistent with clock line spectroscopy. \label{fig:narrow}}
\end{figure}

In order to independently measure low temperatures that we get in the vertical direction, we used clock line spectroscopy in a free space which we  {were able to do} due to the experiments with optical clocks. 
After the capturing of atoms into the 3{rd} MOT (as in Figure \ref{fig:scheme}b), we wait about $20\unit{ms}$ for the magnetic fields to turn off. 
During that time, only $506.2\unit{nm}$-molasses work to restrain atomic cloud from expanding. 
After that, a $1\unit{ms}$-long clock pulse excites atoms along the vertical direction, and we detect the remaining atoms with the probe pulse and a CMOS-camera. 
The result of this experiment is presented in Figure \ref{fig:narrow}a.
The full width at half maximum of this line equals $\Delta\nu_\textrm{clock}=13.7(1.2)\unit{kHz}$. 
In Figure \ref{fig:narrow}b, we can see the result of the time-of-flight experiment, giving  $T = 0.69(6)\unit{\textmu K}$.
This temperature corresponds to $\Delta\nu_\textrm{D}=12(1)\unit{kHz}$ Doppler broadening of the clock transition linewidth.
The similarity of the measured linewidth $\Delta\nu_\textrm{clock}$ and the Doppler broadening $\Delta\nu_\textrm{D}$ assures correct measurements of the temperature of the atoms above and smallness of other residual broadenings, e.g., due to the Zeeman effect.

\subsection{Effect of the 2{nd}-Stage MOT on Recapture Efficiency}
\label{sec:2nd}

\begin{figure}[H]
 
\includegraphics[width=0.98\textwidth]{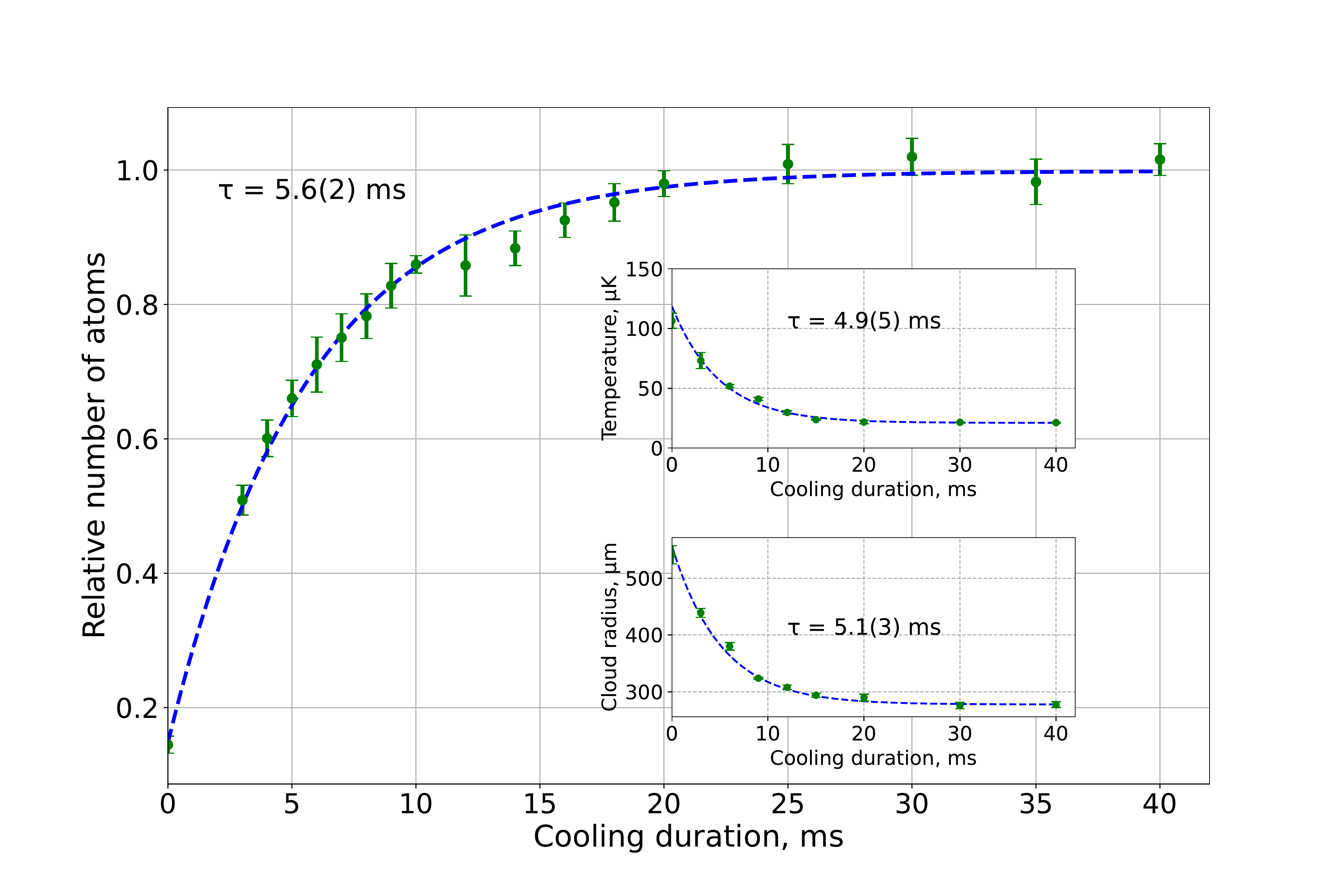}
 
\caption{
Number of atoms recaptured in broad $506.2\unit{nm}$ MOT. Insets: temperature and cloud radius vs. 2{nd}-stage MOT cooling duration.
Dashed lines represent the exponential function fitting, with the characteristic time of the recapturing process being $\approx $$5\unit{ms}$.
}
\label{fig:recapturing}
\end{figure}

While the recapturing efficiency of the broadband MOT $506.2\unit{nm}$ from the 2{nd}-stage MOT is nearly 100\%, we get only $\sim$$15\%$ when loading it directly from the 1{st}-stage MOT. 
In order to understand what impacts this efficiency, we performed the following experiment.
We kept only $410.6\unit{nm}$ and $506.2\unit{nm}$ during the 1{st}-stage MOT (in the experiments described above radiation at $530.7\unit{nm}$ was also present) and varied duration of the 2{nd}-stage MOT (see Figure \ref{fig:scheme}b). 
For each 2{nd}-stage MOT duration $\tau$ the number of atoms recaptured in broadband $506.2\unit{nm}$ MOT is measured (Figure \ref{fig:recapturing}).
Beside this, we also measured the cloud size and the temperature of atoms after the 2{nd}-stage MOT.

All the parameters have similar characteristic time dependence. 
Comparing $\tau=0$ (only the 1{st}-stage MOT) and $\tau>20\unit{ms}$ (steady regime with 2{nd}-stage MOT), the initial cloud size decreases by a factor of 2 and the initial temperature decreases by a factor of 5.
Since we observed almost 7-fold improvement of the number of captured atoms, we conclude that probably both these factors affect the capture efficiency, while the temperature seems to be the dominant one.

It is worth noting that with some optimization of the 1{st}-stage MOT, we could achieve $30\%$ recapture efficiency directly to the broadband $506.2\unit{nm}$ MOT, with the downside of reduced initial number of atoms.

\subsection{Two Atomic Clouds Producing}
\label{sec:double}
Narrow-line cooling can also be used in applications that require spatial control of the atomic cloud.
Some recent works \cite{bothwell2022resolving,zheng2022differential} show great interest in space resolved spectroscopy between separated regions of the atomic cloud or between different clouds in the same vacuum chamber.
During the multistage cooling described above, one can alter the position of the atomic cloud by varying frequency detuning or magnetic field gradient, and thus control the process of recapturing atoms in the optical lattice creating multiple clouds.

In order to achieve two non-intersecting atomic clouds we keep optical lattice and frequency-broadened $506.2\unit{nm}$ radiation on throughout the experiment. 

\begin{figure}[H]
 
\includegraphics[width=13.5cm]{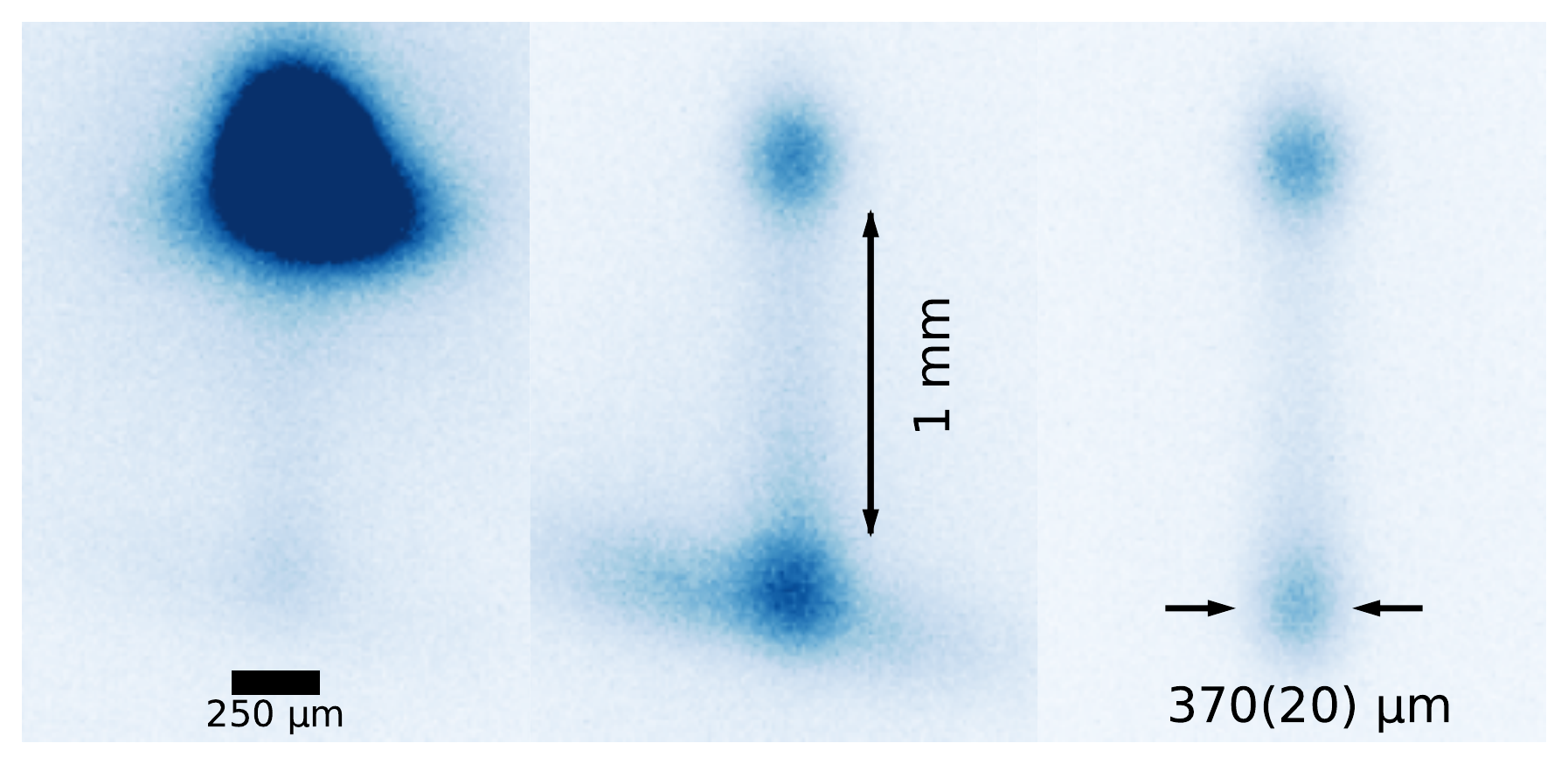}
 
\caption{
Double cloud structure. Three stages of experiment are shown from left to right: 2{nd}-stage MOT, intermediate process of recapturing into 3{rd}-stage MOT, two clouds in optical lattice without any MOT radiation and magnetic fields.}
\label{fig:double}
\end{figure}

First, atomic cloud is formed by $\sim$$50\unit{ms}$-long second stage MOT (see left image on Figure \ref{fig:double}).
Since atoms are already relatively cold ($T\approx20\unit{\textmu K}$), $\sim$$10\%$ becomes captured in the optical lattice at the 2{nd}-stage MOT cloud position. 
After that, $530.7\unit{nm}$ radiation turns off, and anti-Helmholtz magnetic field switches to the 3rd stage cooling regime.
Atoms, which are not trapped into the optical lattice, move to the new position determined by the parameters of the $506.2\unit{nm}$ MOT, as shown in the middle image of Figure \ref{fig:double}.
One can see that a small number of atoms is captured into the optical lattice in between positions of the  2{nd} and 3{rd} MOTs since the atomic cloud was shifting slowly down.
Finally, we switch off $506.2\unit{nm}$ MOT, and after non-trapped atoms fly away, we obtain two separated clouds in the optical lattice (right image in Figure \ref{fig:double}).

Parameters of the frequency-broadened 3rd stage MOT were adjusted so that the resulting atomic clouds contain an equal number of atoms (about 7\% of the initial number of trapped atoms in each cloud) and are spaced several cloud diameters apart (Figure~\ref{fig:double}).
The main characteristics of resulting atomic clouds are mostly determined by the optical lattice.
Note that here the optical lattice was formed by a single reflecting mirror, unlike some of our previous works with enhancement cavity \cite{Golovizin2019, mishin2022effect}. 
Final lifetimes and temperatures of clouds are similar, which enables space-resolved spectroscopy in two non-intersecting atomic clouds and is of interest for future works in this direction.







\section{Discussion}
\label{sec:discuss}

The overall result of this work is extensive experimental analysis of different regimes of a narrow line MOT on $506.2\unit{nm}$ transition in thulium. 
We have shown that we can achieve temperature of the atomic cloud as low as $0.4\unit{\textmu K}$.
When aiming for the highest phase-space density, we were able to reach $\text{PSD} = 3.5\times 10^{-4}$ at $T=0.8\unit{\textmu K}$ when starting with $N=2\times10^6$ atoms in the 1{st}-stage MOT. 
This value is similar to those reported in other experiments with narrow-line MOTs \cite{lu2011strongly, kraft2009bose, aikawa2012bose, stellmer2009bose}.
Implementation of this technique to experiments with thulium with $N=30\times10^6$ \cite{davletov2020machine} or $13\times10^6$ \cite{golovizin2021compact} atoms in the initial MOT, would result in $\text{PSD} \sim 3\times 10^{-3}$. 
This value is 2 orders of magnitude larger than the initial PSD in \cite{davletov2020machine}, that should significantly speed up and increase efficiency of the following evaporative cooling.

The best performance of the narrow-line MOT is achieved using two preliminary stages of laser cooling on 410.6\,nm and 530.7\,nm transitions. 
Meanwhile, we have demonstrated up to 30\% efficiency of the narrow-line MOT loading directly from the 1{st}-stage MOT, which operates on strong 410.6\,nm transition.
This is similar to some experiments with narrow-line MOTs \cite{Degenhardt2005, lu2011strongly}, while less than in other works \cite{katori1999magneto}, depending on atomic species.
Hereby, one can choose between simplicity of the laser setup (i.e., the use of only 2 lasers at 410.6\,nm and 506.2\,nm) and performance, when using 3 stages of laser cooling.

Demonstrated deep laser cooling below $1\unit{\textmu K}$ would significantly improve optical lattice clocks since mostly the ground motional state should be occupied at such temperature and typical depth of optical lattice.
For thulium atoms in the motional ground state we expect clock transition frequency shift from the optical lattice to be at the $10^{-17}$ level with uncertainty at $10^{-18}$, including contribution from the higher-order polarizabilities.

Large sensitivity of $506.2\unit{nm}$ MOT to the parameters of the experiment (like magnetic field gradient, power, and frequency of the cooling beams) can be used  to achieve various advantages dependent on the task. 
One particular example is the use of sensitive 3{rd} MOT for creating double-structured atomic cloud. 
By adjusting cooling parameters, it is possible to get two clouds separated by 1 mm in the same optical lattice. 
Working with two separate but spatially close clouds, it is possible to eliminate a number of inconvenient frequency shifts and noise sources \cite{zheng2022differential, bothwell2022resolving}.

\vspace{6pt}

\authorcontributions{Conceptualization, A.G. and N.K.; data curation, D.T. and A.G.; formal analysis, D.P. and D.T.; funding acquisition, A.G. and N.K.; investigation, D.P., D.T., D.M., M.Y. and A.G.; methodology, D.P., D.T. and D.M.; project administration, K.K. and N.K.; resources, D.K., V.S. and K.K.; software, D.T. and A.G.; supervision, V.S. and N.K.; validation, D.P., D.T., D.M. and A.G.; visualization, D.P. and D.T.; writing---original draft, D.P. and D.T.; writing---review and editing, D.M., A.G. and N.K. All authors have read and agreed to the published version of the manuscript.
}

\funding{This research was funded by Russian Science Foundation grant number 21-72-10108.}

 \dataavailability{The associated experimental data will be available from the authors upon reasonable request.}


 \conflictsofinterest{The authors declare no conflict of interest.}



\abbreviations{Abbreviations}{
The following abbreviations are used in this manuscript:\\

\noindent 
\begin{tabular}{@{}ll}
MOT & Magneto-optical trap\\
BEC & Bose-Einstein condensate\\
PSD & phase-space density\\
AOM & acousto-optic modulator\\
ULE & ultra-low expansion glass \\
\end{tabular}
}

\appendixtitles{yes} 
\appendixstart
\appendix
\section[\appendixname~\thesection]{Laser Frequency Stabilization and Calibration}
\label{sec:append}

 {As a} source of laser radiation at a wavelength of $\lambda = 506.2 \unit{nm}$, we use Toptica Dl pro laser which provides up to $15 \unit{mW}$ optical power with a linewidth less than $100\unit{kHz}$.
The laser frequency is locked to one of the eigenmode of an utrastable cavity made of ultra-low expansion (ULE) glass.
The temperature of ULE cavity was set near its zero thermal expansion point at $T=26.8~^\circ \text{C}$.
The cavity finesse is 47,500 and transmission is $37 \%$ in resonance.  
We stabilized the frequency of the laser to the ULE cavity mode, nearest to the previously investigated transition frequency \cite{provorchenko2021investigation}.



In order to be able to operate with the frequency detuning of the $506.2 \unit{nm}$ laser relative to the exact resonance as well as to compensate the long-term frequency drift of the reference ULE cavity, we perform transition resonance determination at least once a day.
For this, we record absorption spectrum of the cooling $506.2\unit{nm}$ radiation in cloud of thulium atoms, which is formed by the 2{nd}-stage MOT at $530.7\unit{nm}$.

\begin{figure}[H]
 
\includegraphics[width=\textwidth]{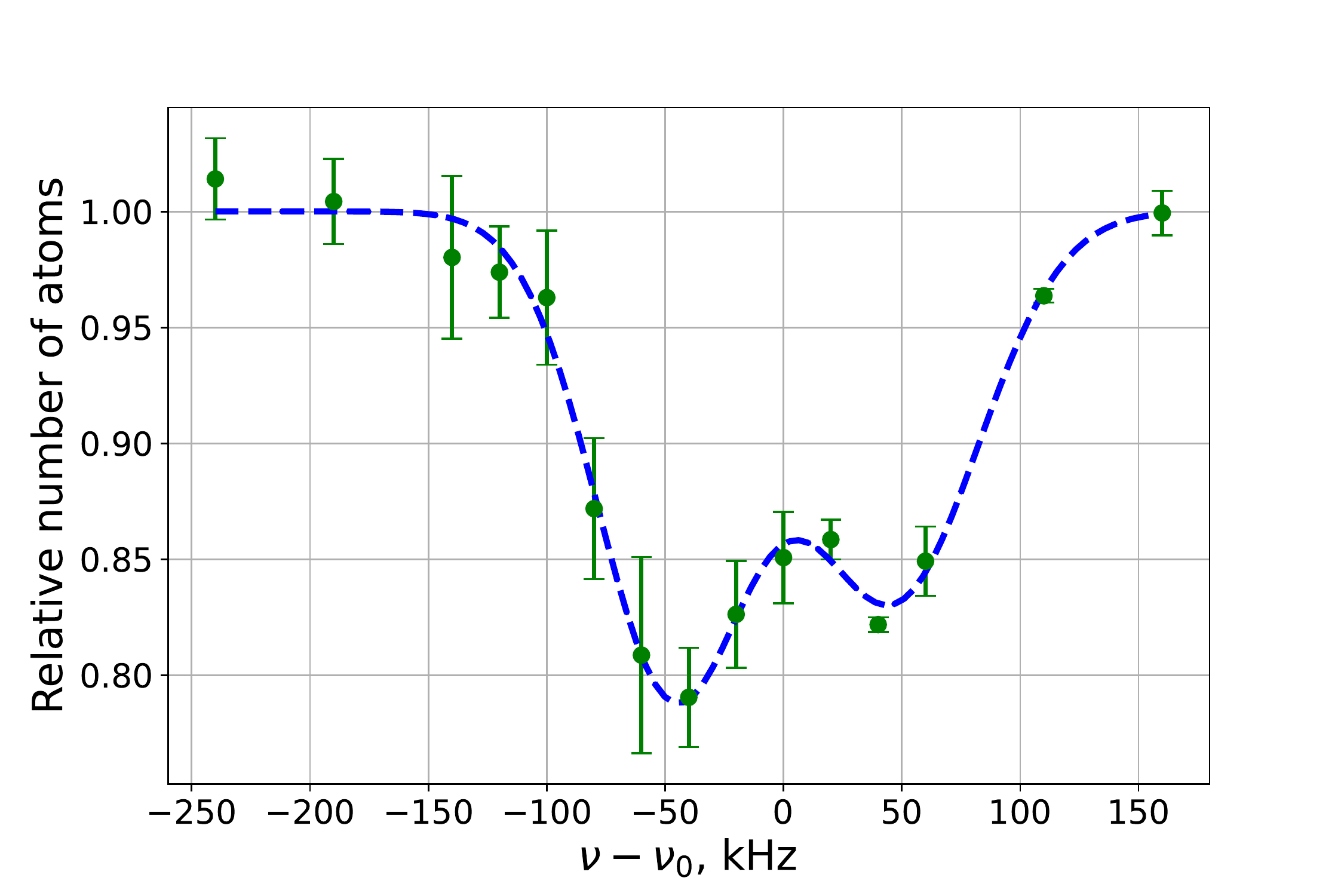}
 
\caption{
Typical $506.2 \unit{nm}$ cooling transition absorption spectrum in free space depending on the AOM frequency $\nu$ detuned from $\nu_0 = 200.985\unit{MHz}$. 
Dashed line shows the results of two Gaussian fit.
The half-sum of their center frequencies was chosen as an origin on the frequency scale in further~experiments.
}
\label{fig:LineCenter}
\end{figure}

After loading of atoms into the MOT, we switched off all the magnetic fields and waited for $30\unit{ms}$ with only $530.7\unit{nm}$ molasses working to prevent the expansion of the atomic cloud during that time.
Then we excited atoms with a $506.2\unit{nm}$ MOT beams and determined the probability of excitation with a probe $410.6\unit{nm}$ beam and a CMOS-camera. 
Scanning the detuning of the $506.2\unit{nm}$ laser, we obtained the spectrum (one of the measurements is shown in Figure \ref{fig:LineCenter}). 

We attribute the double-peak structure to Zeeman shift for different magnetic sublevels of thulium. 
The magnetic field inside the vacuum chamber might be inhomogeneous, and because of large atomic cloud in this experiment, different atoms might be in different magnetic field, and possibly on different magnetic sublevels. Taking into account $g$-factors of the ground and excited states for the  $506.2\unit{nm}$ transition \cite{martin1978atomic}, this structure can be explained by $\approx$$10 \unit{mG}$ residual magnetic field difference on a scale of atomic cloud diameter $\sim$$0.5\unit{mm}$.
The double-peak structure is consistent, and is smoothed out with power broadening of the $506.2\unit{nm}$ transition. 
We choose average frequency between the two peaks as an origin on the frequency scale because of the symmetry of the experiment. This corresponds also to the line center of the power-broadened spectrum.

From day-to-day calibration we observe linear cavity drift of $\approx$$30\unit{kHz/day}$. 
On a typical timescale of our experiment of $\sim$$4\unit{h}$, this results in $\approx$$5\unit{kHz}$ frequency shift which is on the order of the transition natural linewidth ($\gamma=7.8\unit{kHz}$). 
As we typically work at the frequency detuning of $-10\gamma$, the laser frequency drift does not affect the experiments.
We note, that in the future we plan to automatically compensate this drift with an acousto-optic modulator in front of the reference cavity.


\begin{adjustwidth}{-\extralength}{0cm}

\reftitle{References}



\PublishersNote{}
\end{adjustwidth}
\end{document}